\documentclass[letterpaper,preprint,rsi,superscriptaddress,floatfix]{revtex4}
\usepackage[german,swedish,english]{babel}
\usepackage[dvips]{graphicx}
\usepackage{amssymb}
\usepackage{amsmath}
\usepackage{color}

\begin{document}

\date{\today}

\title{A single-solenoid pulsed-magnet system for single-crystal scattering studies}
\author{Zahirul Islam}
\affiliation{X-Ray Science Division, Advanced Photon Source, Argonne National Laboratory, 9700 S. Cass Ave., Argonne IL 60439\\}
\author{Dana Capatina}
\affiliation{APS Engineering Support Division, Advanced Photon Source, Argonne National Laboratory, 9700 S. Cass Ave., Argonne IL 60439\\}
\author{Jacob P. C. Ruff}
\author{Ritesh K. Das}
\affiliation{X-Ray Science Division, Advanced Photon Source, Argonne National Laboratory, 9700 S. Cass Ave., Argonne IL 60439\\}
\author{Emil Trakhtenberg}
\affiliation{APS Engineering Support Division, Advanced Photon Source, Argonne National Laboratory, 9700 S. Cass Ave., Argonne IL 60439\\}
\author{Hiroyuki Nojiri}
\author{Yasuo Narumi}
\affiliation{Institute for Materials Research, Tohoku University, Sendai, Japan\\}
\author{Ulrich Welp}
\affiliation{Materials Science Division,  Argonne National Laboratory, 9700 S. Cass Ave., Argonne IL 60439\\}
\author{Paul C. Canfield}
\affiliation{Department of Physics and Astronomy, Ames Laboratory, Iowa State University, Ames IA, 50010}

\begin{abstract}
We present a pulsed-magnet system that enables x-ray single-crystal diffraction in addition to powder and spectroscopic studies with the magnetic field applied on or close to the scattering plane. The apparatus consists of a single large-bore solenoid, cooled by liquid nitrogen. A second independent closed-cycle cryostat is used for cooling samples near liquid helium temperatures. Pulsed magnetic fields close to $\sim 30$ T with a zero-to-peak-field rise time of $\sim$2.9 ms are generated by discharging a 40 kJ capacitor bank into the magnet coil. The unique characteristic of this instrument is the preservation of maximum scattering angle ($\sim 23.6^\circ$) on the entrance and exit sides of the magnet bore by virtue of a novel double-funnel insert. This instrument will facilitate x-ray diffraction and spectroscopic studies that are impractical, if not impossible, to perform using split-pair and narrow-opening solenoid magnets and offers a practical solution for preserving optical access in future higher-field pulsed magnets.
\end{abstract}

\maketitle

\section{Introduction}
The use of high-field pulsed magnets with various x-ray and neutron techniques\cite{minicoil1,minicoil2,xdiff1,xdiff2,xdiff3,xdiff4,xdiff5,ruff10,rsi09,pow1,pow2,PvdL,minicoil3,minicoil4,val,xmcd,xmcd2,xres,neut1,neut2,neut3} is being widely explored in order to develop a fundamental understanding of material properties in extreme magnetic fields \cite{hifield1,hifield2,nrc,cmmp2010,xtreme,jpn-100T}. A number of high field (30-40 T) pulsed magnets in two common geometries discussed below are in operation at third-generation synchrotron sources. These include small-bore and large-bore coils that generate pulsed fields with a duration of $\sim$1 to 5 ms in split-pair as well as single solenoid geometries, utilizing capacitor banks discharging anywhere from $\sim$2 kJ to more than  $\sim 200$ kJ of energy \cite{minicoil1,minicoil2,minicoil3,minicoil4,neut1,val}. The primary advantage of a split-pair geometry is optical access, albeit with considerable mechanical design challenges to keep the two magnets from collapsing on each other. On the other hand, a solenoid offers the largest pulsed magnetic fields at the center albeit with a much more limited optical access. In order to reach liquid helium temperatures for the sample, however, complex cryostat designs need o be incorporated which further reduce optical access in this geometry. As a result, pulsed field studies using single-solenoid magnets have, thus far, been confined to spectroscopic and powder diffraction experiments.

Magnetic-field induced effects are in general anisotropic, and in many cases novel states only appear when the field is applied along certain crystallographic axis. Recent diffraction studies of Jahn-Teller (JT) distortions \cite{jt1,jt2,pow1,pow2}, structural transitions in oxides and multiferroics \cite{xdiff1,xdiff2,xdiff3,xdiff4,lufeo}, and magnetostriction (MS) in rare-earth compounds \cite{xdiff5,ruff10}, highlight the need to determine structural responses in materials in high magnetic fields by employing single-crystal diffraction (SXD) techniques. However, SXD often needs large scattering angles. Therefore, such structural studies in fields reaching 60 T and beyond would require the use of single-solenoid pulsed magnets with substantial optical access. In this article we describe a single-solenoid magnet with a mechanical design and dual-cryostat scheme that preserves optical access to the sample. It can generate pulsed fields up to $\sim$30 T with total duration of $\sim$6 ms every $\sim$8 min. using a modest 40 kJ of energy. A maximum of $\sim23.6^\circ$ of scattering angle is accessible to study samples at or below liquid helium temperatures. Pulsed-field induced JT effects in TbVO$_4$ and single-crystal diffraction on YBa$_2$Cu$_3$O$_{6+x}$ (YBCO) were observed to illustrate the performance of this instrument.
 
\section{Experimental details}
The pulsed magnet system consists of (1) a liquid nitrogen (LN$_2$) bath cryostat for the solenoid, (2) a closed-cycle cryostat for sample, (3)  a 40 kJ capacitor bank, and (4) portable control and data acquisition equipment rack. In what follows, we describe the arrangement of the two cryostats, a thermal analysis for cooling samples, the characterization of generated pulsed fields, and x-ray studies used to calibrate the system.

\subsection{Dual cryostat}
\begin{figure}[ht]
\includegraphics[width=0.8\textwidth]{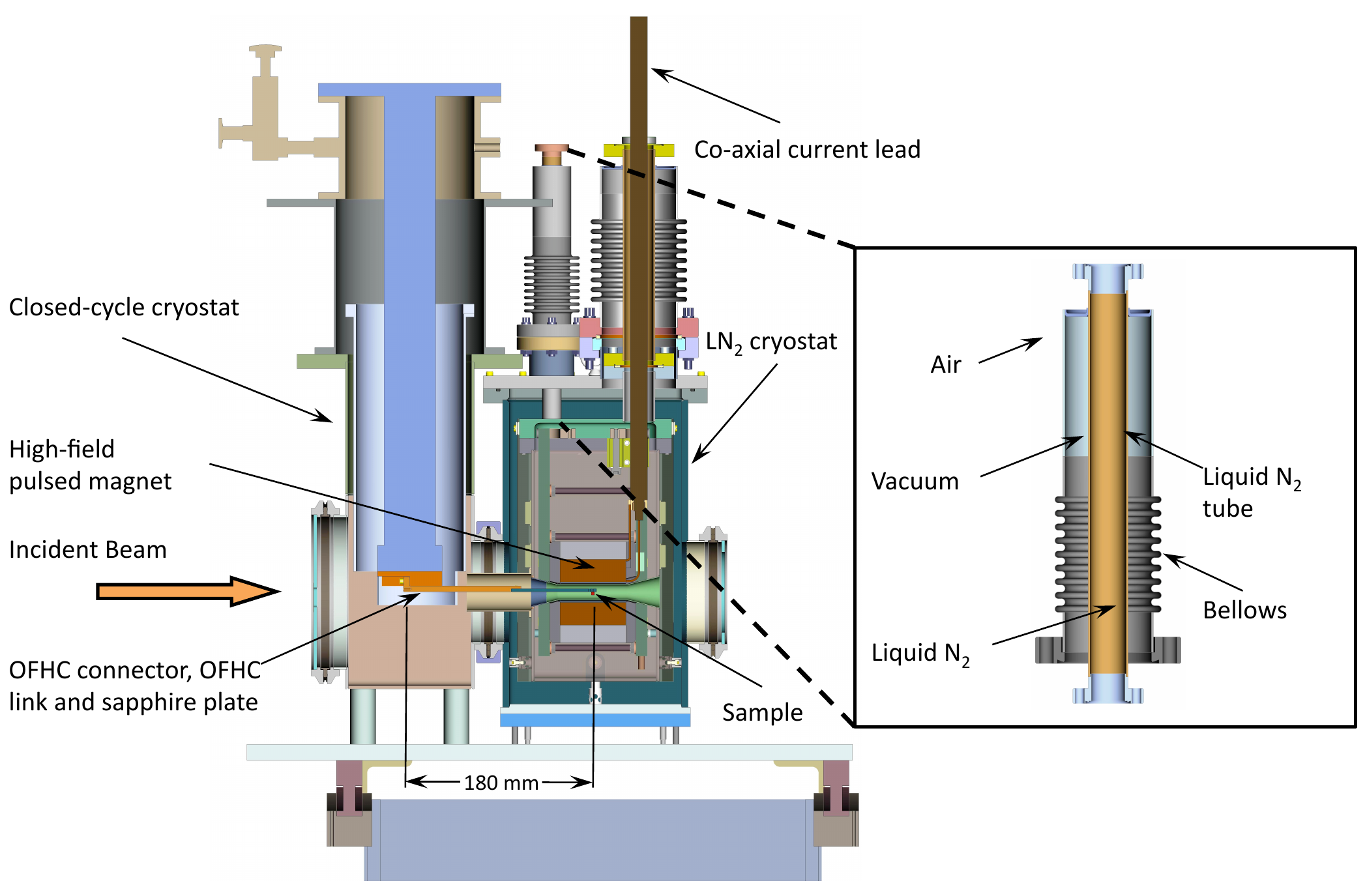}
\caption{Dual-cryostat scheme. The LN$_2$ cryostat on the right contains the pulsed magnet. The closed-cycle cryostat on the left is for the sample. They are coupled via a vacuum shroud with X-ray windows on either side. The cryostats are kept in the vertical orientation with scattering plane being horizontal. The inset on the right shows how the LN$_2$ ports are nested through the vacuum jacket using bellows.}
\label{2cryo}
\end{figure}

\begin{figure}[ht]
\includegraphics[width=0.7\textwidth]{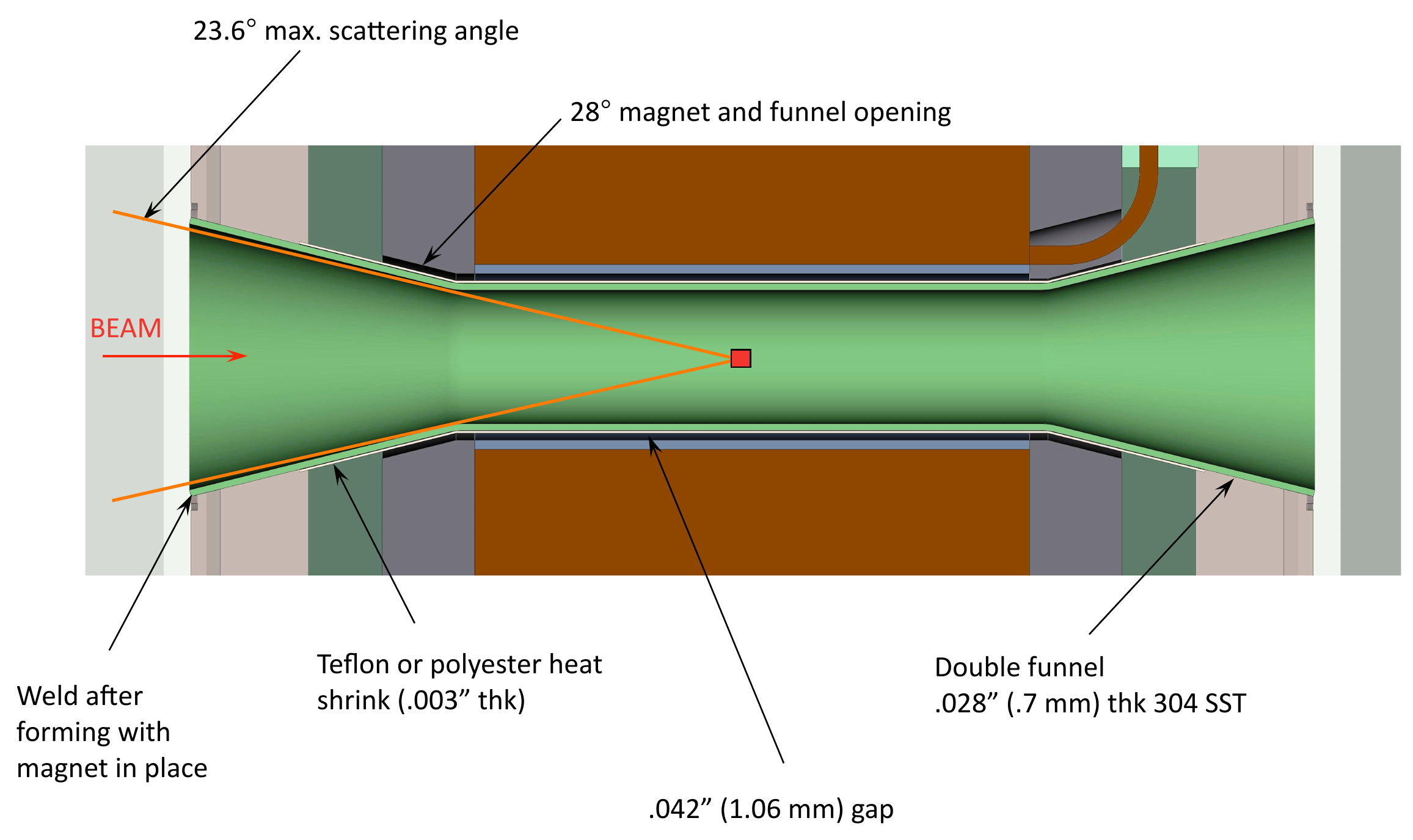}
\caption{Details of the double funnel insert. Note the small gap between the funnel wall and inside wall of the magnet bore which is sufficient to keep them apart at LN$_2$ temperature.}
\label{2funnel}
\end{figure}

In order to independently cool the solenoid and the sample we have developed separate cryostats (Fig\ \ref{2cryo}) for cooling the magnet and sample respectively. A commercial closed-cycle refrigerator (1 W of cooling power at 4 K with 1.4 K base temperature without a heat load from Cryo Industries of America) with large diameter cold-finger was chosen to cool the sample, while for solenoid cooling we have designed and built a LN$_2$ bath cryostat. These two cryostats are adjacent to each other with their vacuum shrouds coupled via flanges making a common vacuum space. At the core of the LN$_2$ bath cryostat is a ``double-funnel'' (see Fig.\ \ref{2funnel}) which goes through the magnet bore\cite{medsi} and separates the vacuum space from the LN$_2$ in the bath. It is made of stainless steel thin enough ($\sim$0.7 mm) to allow a complete magnetic flux penetration during field pulses (see below). It has two flared ends one of which was formed after insertion through the magnet bore. Both ends were welded to the outside wall of the cryogen vessel. The sample space inside the double-funnel is connected to the cryostat vacuum space and it is separated from the LN$_2$ by the double-funnel wall. This funnel preserves most of the optical access allowed by the magnet bore as depicted in Fig.\ \ref{2funnel}. The kapton insulation of the innermost layer of the coil is exposed for efficient direct-contact cooling with LN$_2$ filling the gap between the funnel and the magnet bore. Note that the outer wall (LN$_2$ side) of the funnel is wrapped with Teflon tape providing additional insulation between exposed metal and the inner layer of the coil. The overall weight of the system is $\sim 148$ kg. The footprint of the system is $\sim 375$ mm$ \times \sim 425$ mm with a height of $\sim 700 $ mm (excluding the length of the co-axial current leads above the sample cryostat in Fig.\ \ref{2cryo}).

X-rays enter the dual cryostat from the sample cryostat side (left in Fig.\ \ref{2cryo} and Fig.\ \ref{2funnel}), defined to be the upstream side. LN$_2$ is then on the downstream side with respect to the sample cryostat. The sample mount consists of an oxygen-free high thermal conductivity (OFHC) copper connector, an OFHC copper link and a sapphire plate. The copper connector mounts on the cold finger, while the sample is glued just below the end of the sapphire plate, at the center of the coil. For good thermal contact between the connecting parts, compressible indium foil can be used with adequate pressure. An intermediate radiation shield around the cold finger can be attached to both the sample-cryostat radiation shield ($\sim$30 K) and to the LN$_2$ tank ($\sim$77 K) in order to minimize the radiation heating from the room-temperature ambient. With such a configuration (Fig.\ \ref{2cryo}) the distance between the vertical axis of the sample cryostat and sample at the center of the magnet coil is 180 mm.

During operation the LN$_2$ magnet cryostat is kept under positive pressure to prevent condensation of any kind to develop, in particular on the high-field coil. It has one fill port connected to a large supply dewar through a solenoid valve. There are two exhaust ports each of which is connected to a 5 m long 50 mm diameter bellows terminated by check valves. These bellows act as a buffer volume against any sudden pressure build ups. One of them has a 9-11.5 psig burst disk and acts as the relief line. The system is pumped out and purged with ultra-pure dry air through ports on the bellows to eliminate moisture before filling the vessel with LN$_2$. A nitrogen level meter constantly checks the amount of nitrogen in the vessel and controls the solenoid valve to allow LN$_2$ from the dewar to fill the vessel up to a preset level. 

We note that in its current state the cryostat is designed for continuous operation for a few days with one or two samples attached to the end of the sample mount. However, if needed samples are changed without completely warming up the system. In order to change samples the LN$_2$ bath is first emptied and kept under flowing dry air while the sample cryostat is warmed up to room temperature. After $\sim$8 hours the sample vacuum is broken by flowing inter gas into the system. The sample mount is removed while flowing inert gas through the system for changing sample, which takes less than 2 hours.

\subsection{Thermal Analysis}
\begin{figure}[ht]
\includegraphics[width=1.0\textwidth]{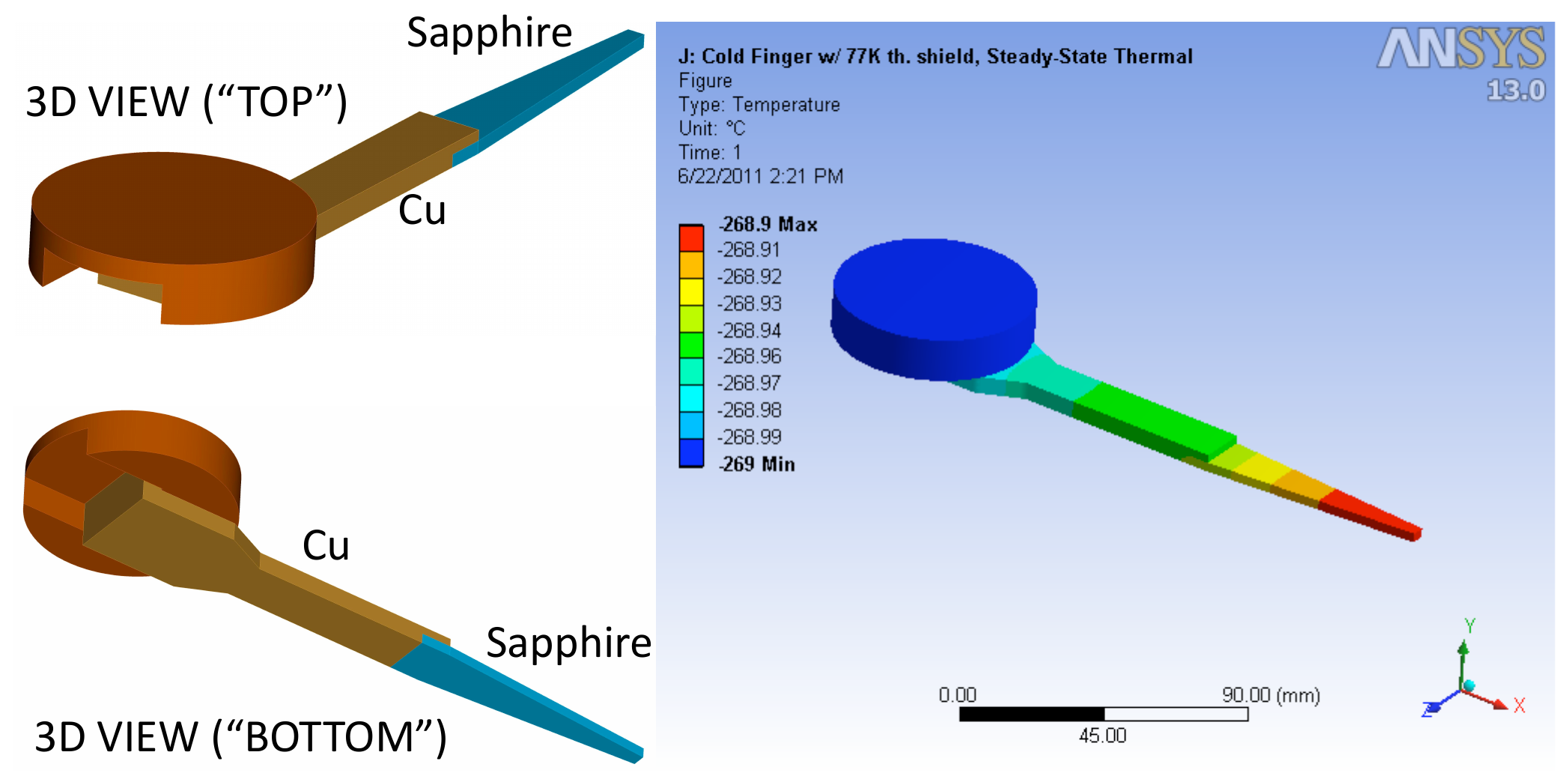}
\caption{Left: Isometric (3D) views of the sample mount from two perspectives. Right: Temperature distribution across the length of the sample mount with a radiation shield at $\sim$77 K.}
\label{coldfin}
\end{figure}

Since the sample is at a substantial distance downstream from the cold finger, a finite element thermal analysis \cite{ansys} was carried out in order to simulate the temperature distribution along the sapphire plate and the heating by thermal radiation. The model consisting of an OFHC copper connector, the OFHC copper link and the sapphire plate was built in Pro Engineer package (PTC) and is shown in Fig.\ \ref{coldfin}. Features that are not relevant for this analysis such as mounting holes, chamfers and fillets were omitted to simplify computation, as they do not affect the analysis results. The material property needed for the steady-state thermal analysis is thermal conductivity ($\kappa$) of OFHC copper and sapphire at low temperature. A conservative value for thermal conductance across both copper-to-copper and copper-to-sapphire interfaces, which emulates a poor contact ({\it e.g.} using indium foil with low contact pressure), to ensure that the proposed scheme is reasonable for reaching the base temperature was used. A constant temperature of 4 K was applied on the OFHC copper connector top surface, and radiation to ambient was applied on all the other surfaces. A conservative emissivity coefficient of 0.5 was used in the analysis. 

Three cases have been analyzed: (1) No radiation shield; (2)	30 K radiation shield; and (3) 77 K radiation shield. The primary  difference between the cases lies in the ambient temperature surrounding the copper link and the sapphire plate. The intermediate radiation shield attached to the sample cryostat radiation shield (case (2)) reaches around the copper link and the sapphire plate. The intermediate radiation shield attached to the LN$_2$ tank (case (3)) goes from the tank up to the sample cryostat. 

Without using a radiation shield (Case (1)) the system would experience about 1 W of radiative heating, while use of a radiation shield decreases the radiation heat to about 43 mW for the 77 K option or to about 2 mW for the 30 K option.  The calculated temperature difference between the (4 K) cold finger and the downstream end of the sapphire plate was 1.1 K without a shield and 0.1 K with a 77 K shield. Nevertheless, to keep a constant temperature of 4 K at the cold finger, without using a cold finger radiation shield, the sample cryostat will have to overcome a radiative heating of 1 W, independent of the quality of the interface contact. We conclude that a sample base temperature at or below 4 K should be achievable using the designed mount and radiation shield.

The temperature distribution results for case (3) are shown in Fig.\ \ref{coldfin}. In case (2) the base temperature is slightly lower. However, implementation of a 30K radiation shield surrounding the sapphire plate was impractical. Instead we have installed a pair of half cylinders between the LN$_2$ tank and the cryostat without interfering with the optical access, which is surrounded by the 77 K shield. Although there is a 77 K radiation shield from tank reaching the sample cryostat we note that there is a substantial solid angle of the downstream end of the double-funnel that was kept open for x-rays to pass through allowing some room-temperature radiation heat to reach the sample. Despite this additional heat a base temperature of 6K was reached at the sample position. We note that there are mounting holes on LN$_2$-bath wall to attach coated sapphire (or aluminum foils in the case of high-energy x-rays) plates to close the downstream end of the double-funnel, which will be implemented to improve radiation shielding.

\subsection{Pulsed magnetic fields}
\begin{figure}[ht]
\includegraphics[width=0.5\textwidth]{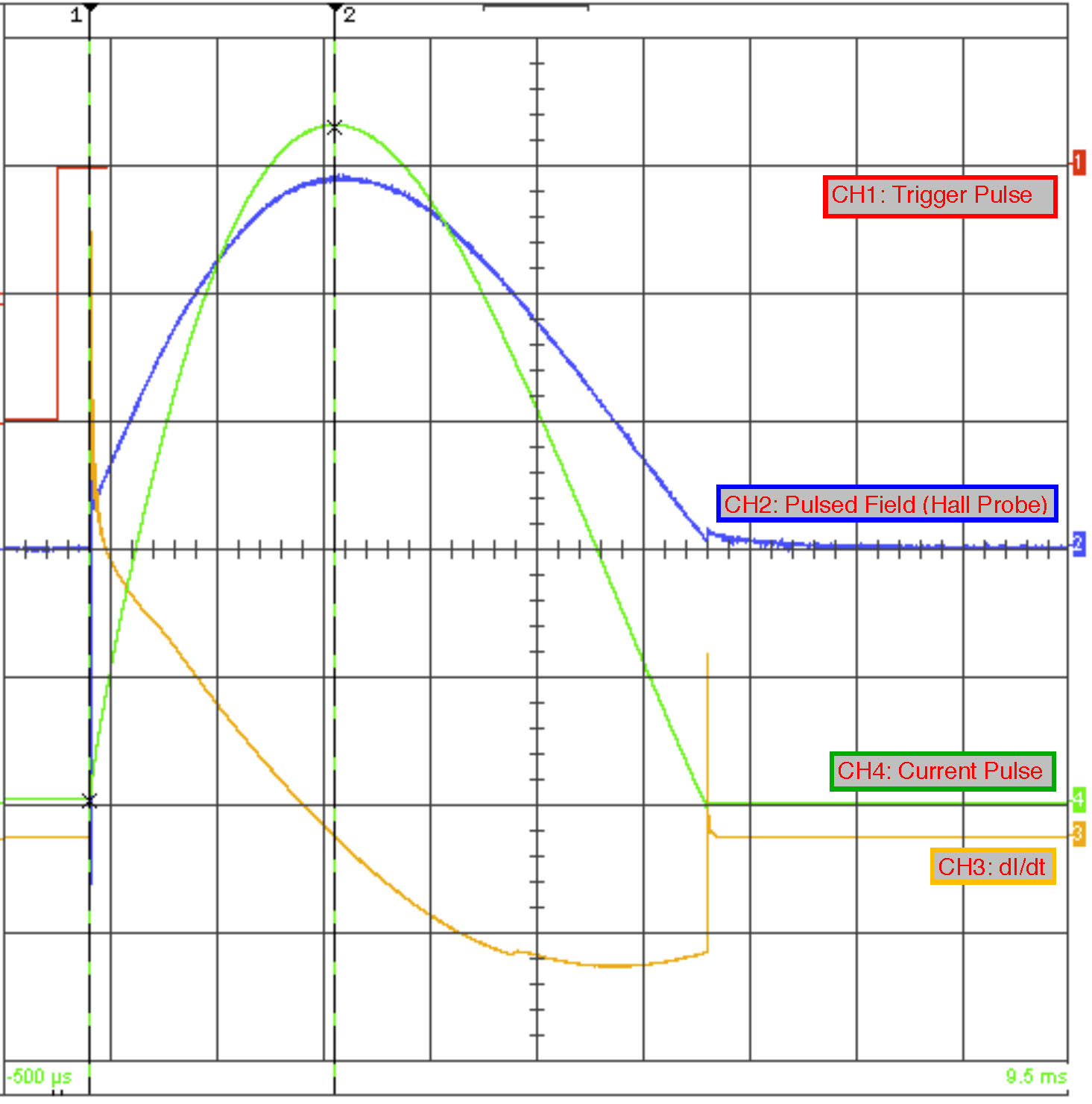}
\caption{A typical screenshot of DSO during pulsed field generation. Pulsed field (CH2) and current (CH4) profiles measured using a Hall probe and current transformer with peaks corresponding to $\sim$29.2 T and 10.6 kA, respectively. $\frac{dI}{dt}$ from the capacitor bank and the primary trigger pulse are captured in CH3 and CH1. (DSO: Vertical scale is 2 V/div.)}
\label{dso}
\end{figure}

The solenoid in this instrument was designed and built at Tohoku University. It is wound using high-tensile-strength CuAg wires (2X3 mm$^2$) with kapton insulation\cite{cuag}. The coil has a 18 mm of inner diameter and 65 mm of outer diameter, respectively. It is encased in maraging steel with a flange on either end with an overall length and diameter of $\sim$100 mm each. The coil has a long coaxial lead constructed of brass rod as the central conductor and brass tube as the outer conductor, respectively. Pulsed magnetic fields are generated by a capacitive discharge through the magnet coil. At the APS the coil is used with a 40 kJ (3000 V, 9 mF) capacitor bank \cite{metis, rsi09}, which can be fully  charged in less than 25 s and discharged using a digital trigger (Fig.\ \ref{dso}). The system can generate half-sine and stretched half-sine with extended decay current pulses with the switching capability to reverse current direction in both modes. The bank has automatic timeouts to safely discharge through internal dump resistors. Furthermore, there are interlocks to prevent accidental discharge or inadvertent access attempts. The control and data acquisition rack (described in \cite{rsi09}) primarily consists of a digital delay generator, 10 MHz digital storage scope (DSO), and a multi-channel scaler. A Hall probe (AREPOC) was axially aligned inside the bore  in order to measure the magnetic fields directly for different charging voltages while the current through the coil was measured using a high-precision current transformer (Pearson Model 1423). 

\begin{figure}[ht]
\includegraphics[width=0.8\textwidth]{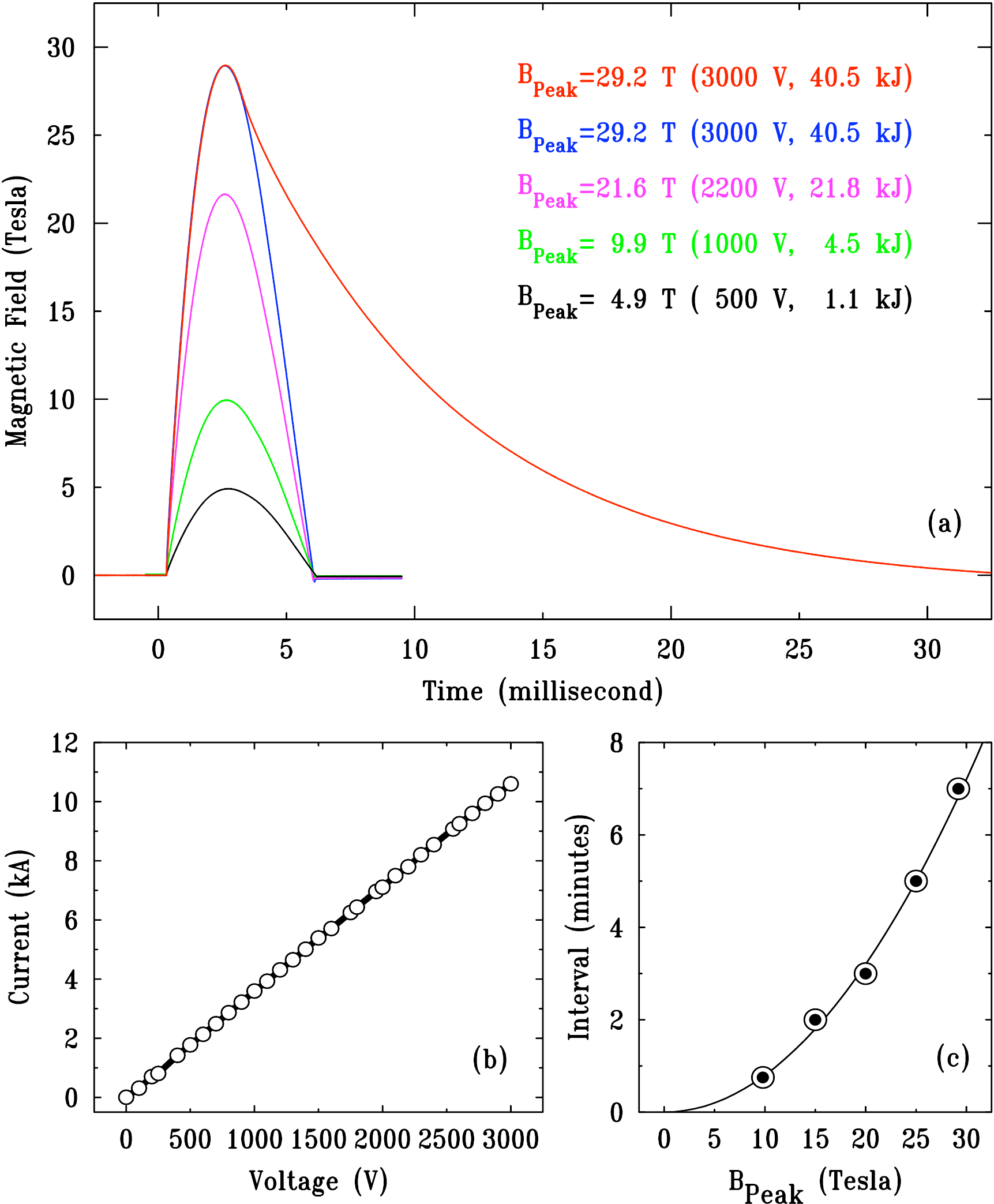}
\caption{(a) Field profiles for different values of the charging voltage. For the highest field profile with an extended decay is also shown. (b)  I$_{Coil}$-V relationship. (c) Minimum time interval between field pulses with B$_{Peak}$.}
\label{prof}
\end{figure}

A set of typical data of transients such as current through the coil (I$_{Coil}$), central field value, etc., during field pulse captured using the DSO is displayed in Fig.\ \ref{dso}. With 40 kJ (3000 V) and the capacitor bank configured to half-sine mode a peak field of above 29 T (corresponding peak current $\sim$10.6 kA) at the bore center is generated. The start of the field pulse as indicated by dI/dt occurs at a fixed delay with respect to the TTL trigger pulse. A set of current pulses were recorded as a function of charging voltage of the capacitor bank as shown in Fig.\ \ref{prof}. The peak current shows a linear behavior with charging voltage. These data allow the determination of the I$_{Coil}$-V relationship for our system and a coil constant of $\sim2.75$ T/kA. Since measured field values using the Hall probe are uniquely determined by the current, field values can be inferred from measured currents without the need for a Hall probe (or a pick-up coil) during the experiment. We note that the coil has been engineered to withstand fields up to 55 T, which is well above the maximum obtainable field using 40 kJ. So, one can expect the coil to have a very long life for routine operations at or below 30 T.

To determine the minimum time needed for the coil to cool down after a pulse we have systematically generated pulses at a fixed charging voltage and varied intervals between successive pulses. By comparing the peak current and pulse shape we determined the minimum cool down time for a given charging voltage. These measurements yielded a repetition rate curve which is shown in Fig.\ \ref{prof}. The wait time increases quadratically with peak magnetic field with a minimum of $\sim 7$ min for $\sim 30$ T. A quadratic behavior implies Joule heating which scales with the $RI^2$, where $I$ is the current through the coil and  $R$ is the total resistance including $\sim 25$ m$\Omega$ of the coil at LN$_2$ temperature.  

\subsection{Jahn-Teller effects in TbVO$_4$}
\begin{figure}[ht]
\includegraphics[width=0.8\textwidth]{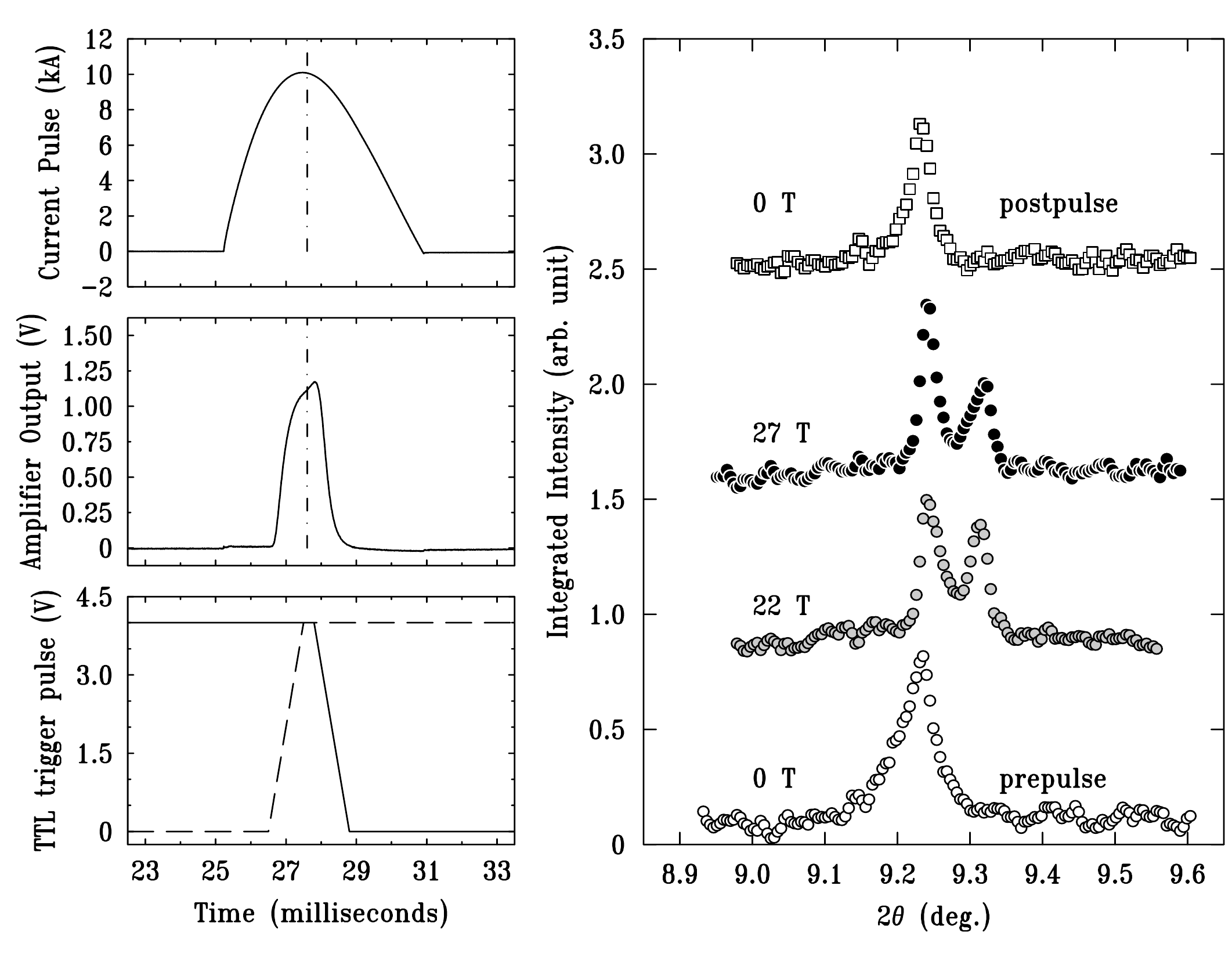}
\caption{Left: Area detector was exposed during a time window (middle panel shows ion chamber output proportional to incident flux) centered on the peak of the pulsed field (top panel). Two shutters were synchronized using TTL pulses to precisely define the exposure window (bottom panel). Right: Jahn-Teller splitting of (220)$_T$ Bragg peak captured at pulsed field maximum. Data were collected above $T_Q$ and shifted vertically for clarity.}
\label{jt}
\end{figure}

In order to demonstrate the use of the pulsed magnet instrument we have performed diffraction studies of field effects on JT distortions in TbVO$_4$ compound\cite{jt1,jt2}. Flux-grown crystals were ground into a fine powder. The powder was mixed with GE varnish in a circular groove on a thin sapphire plate attached to the end of the sample mount (Fig.\ \ref{coldfin}). X-ray diffraction studies were carried out on the APS 6-ID-B beamline. A Si(111) monochromator was substantially detuned in order to select 30 keV photons and suppress higher-order harmonics in the beam. Ideally, one would collect time-resolved diffraction patterns to observe the entire field dependence within a few milliseconds similar to measurements on Tb$_2$Tii$_2$O$_7$ using fast (capable of a full-frame readout at or above 20 kHz) strip detectors\cite{strip}. However, large two-dimensional detectors with such a fast read-out time suitable for collecting powder-diffraction data are not available. We have used an image plate (MAR345) for collecting diffraction data. A pair of fast shutters were synchronized so that the detector was only exposed for $\sim$1 ms with the exposure centered on the peak of the pulsed field (Fig.\ \ref{jt}, left panels)\cite{uniblitz}.  Each shutter has a 6 mm diameter aperture with a minimum exposure time of $> 5 ms$. However, two of them are synchronized so that a much smaller aperture is created to allow a much shorter exposure time. With this scheme by changing the relative delay between the shutters the exposure time can be varied as well. We note that we have measured opening and closing times for each shutter which remained constant throughout the course of the measurements. Each of the shutters made of a Pt-Ir alloy was absorptive enough for 30 keV photons that during the course of the synchronization transmitted beam through the shutters outside the exposure window was negligible. As a result a single exposure was sufficient for collecting clean diffraction patterns. The incident beam transient profile was convolved with measurements to determine precisely the range of fields traversed during data collection. 

TbVO$_4$ undergoes a tetragonal-to-orthorhombic structural phase transition due to cooperative JT distortions at $T_Q\sim$33 K. These distortions are susceptible to magnetic fields, which have been studied in pulsed fields at ESRF \cite{pow1,pow2}. Note that JT splitting manifests itself in some powder lines and not in others. We observed splitting of several Bragg peaks collected at a temperature of $\sim$35 K which is above $T_Q$ consistent with previous work\cite{pow1,pow2}. Fig.\ \ref{jt} (right panel) shows the radially integrated (220)$_T$ Bragg peak (indexed in the tetragonal phase) at 0 T, 22 T, and 27 T, respectively. The (220)$_T$ peak, which was not measured in earlier studies\cite{pow1,pow2}, is expected to show a large magneto-elastic splitting into orthorhombic (400) and (040). According to our data \cite{celath} the peak splits at 22 T by a large amount with their relative intensity changing at higher fields. Furthermore there is a clear magneto-strictive shift of the orthorhombic lattice parameters which manifest as higher Bragg angles for both (400) and (040) peaks relative to that of (220)$_T$.

\subsection{Single-Crystal Diffraction}
\begin{figure}[ht]
\includegraphics[width=0.8\textwidth]{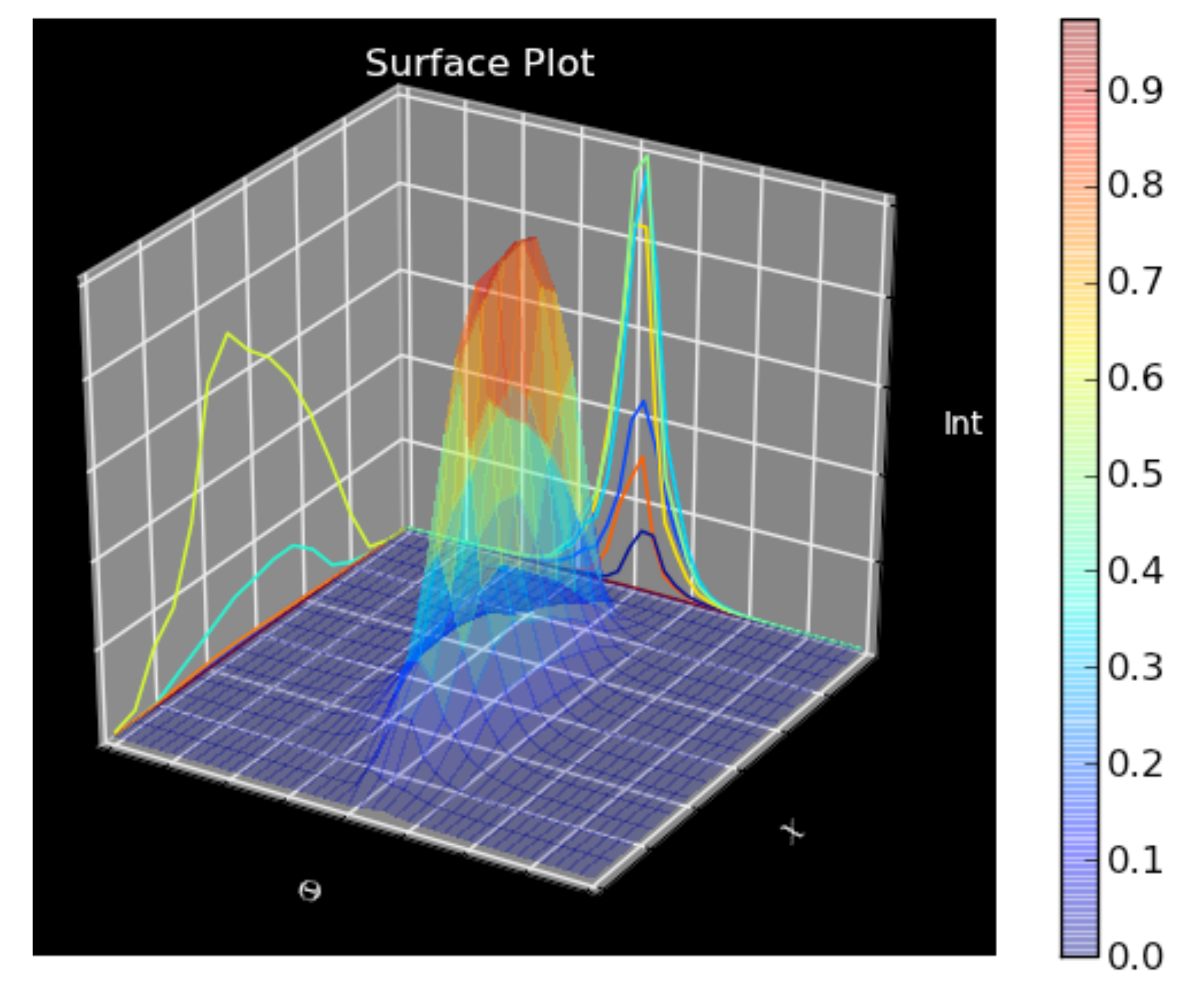}
\caption{Measurements of (4,0,0) Bragg-peak angular widths in two directions show a well-defined and sharp grain. The FWHM in the transverse direction ({\it i.e.} $\theta$) is $\sim0.025^\circ$. Ranges of scans is $0.4^\circ$ and $ 0.6^\circ$ in $\theta$ and $\chi$, respectively.}
\label{mosaic}
\end{figure}

\begin{figure}[ht]
\includegraphics[width=0.8\textwidth]{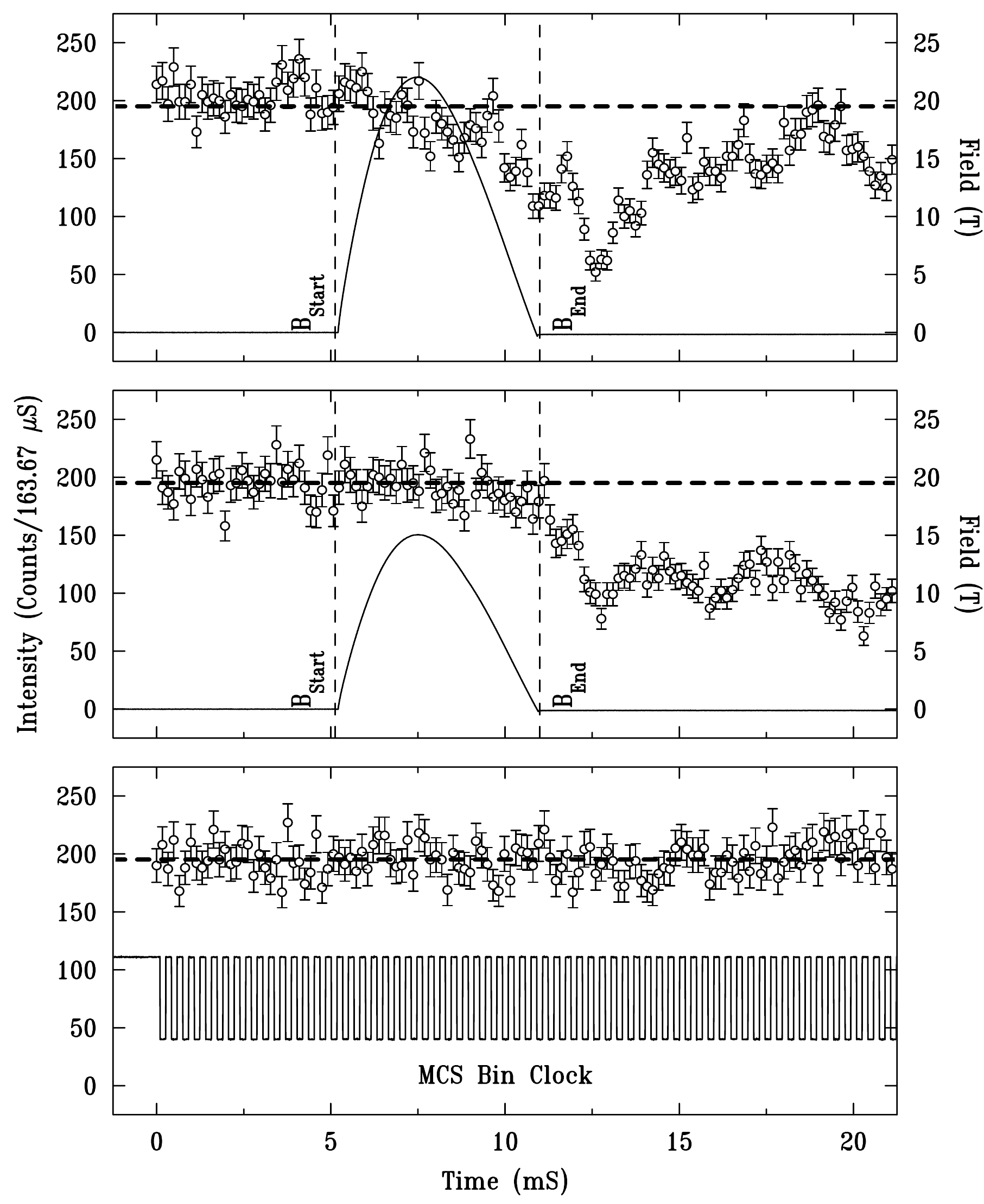}
\caption{SXD data on (4,0,0) Bragg peak of YBCO as a function of time. Data were binned (see bottom panel) by an MCS. Bottom: Bragg-peak intensity during no field pulse. Middle: Intensity oscillations due to vibrations from a $\sim$ 16 T pulse manifest after the pulse. Top: Intensity near the end of the pulse starts to get perturbed by a 22 T pulse.}
\label{ybco}
\end{figure}

The observation of JT effects with the present pulsed magnet system using powder diffraction technique is encouraging, While in TbVO$_4$ it was possible to possible to study magnetic-field induced effects on powdered sample, one needs to bear in mind that such effects are in general anisotropic, and in many cases novel states only appear when the field is applied along certain crystallographic axis. However, the study of single crystals using diffraction methods in pulsed magnetic fields is challenging. During pulsed field generation vibrations from the coil can propagate to the sample and misalign it to compromise the measurements at high fields. We have chosen to study a single-crystal sample of an underdoped YBCO superconductor (SC) to assess the degree to which SXD is susceptible to vibrations. Since a SC below its superconducting transition temperature naturally experiences strong repulsive forces due to its diamagnetic nature, this constitutes a stringent test of vibrational issues. A rectangular single-crystal sample was stress de-twinned and annealed.  It was attached to a thin sapphire plate, which was secured at the end of the sample mount.  The crystal was aligned with the {\bf c}-axis along the magnetic field and its orthorhombic {\bf a}-axis in the horizontal plane. SXD measurements were carried out on the 4-ID-D beamline using 36 keV x-rays. We used a LaCl$_3$ scintillator detector and multi-channel scaler (MCS) in order to collect time-resolved scattered intensity during the field pulse. The crystal was of high-quality as revealed by measurements of intensity as a function of Bragg angle ($\theta$) and tilt angle ($\chi$) with respect to the scattering plane, respectively. Fig.\ \ref{mosaic} shows a surface plot of (4, 0, 0)-peak intensity. There is a single and well-defined peak, which is sharper in $\theta$ with a mosaic width of $\Delta\theta\sim 0.025^\circ$. In Fig.\ \ref{ybco} intensity of the (4, 0, 0) Bragg peak (2$\theta _B\sim20.832^\circ$, FWHM$\approx$0.0041 {\AA$^{-1}$}) is shown as a function of time for two different field pulses (top panels) along with that for no field pulses (bottom panel). For peak fields of $\sim$16 T vibrations cause the Bragg peak intensity to oscillate, although not until long after the pulse. At $\sim$22 T the vibrations start to affect the intensity near the end of the pulse. So, one can measure SXD data throughout the full pulse, as long as peak fields are at or below 22 T. However, vibrations from pulses with peak fields exceeding 22 T start to affect the intensity during the falling half of the pulse. If we assume that the loss of intensity is entirely due to angular motion of the sample then a $\sim50\%$ reduction implies $\sim0.01^\circ$ amplitude for the vibration. Even though the measurements during the falling half of the pulse are not possible for peak fields near $\sim$30 T, it may still be possible to collect field dependence of the SXD data during the rising half of the pulse as long as crystal mosaic is larger than $\sim0.025^\circ$. We note that our instrument is designed to incorporate weak links between the magnet and sample cryostats as well as vibration isolation pads. Further quantitative determination of vibrations and implementation of such links and pads, which should alleviate, if not eliminate, vibration issues from our system, are being planned.

\section{Concluding Remarks}
We have developed a pulsed-field magnet instrument with unique features for x-ray studies at third-generation synchrotron radiation sources. Fields near  $\sim$30 Tesla can be generated repeatedly using a modest 40 kJ of energy. Two independent cryostats cool the magnet coil (using liquid nitrogen) and sample (using a closed-cycle refrigerator), respectively. LN$_2$ cooling allows a repetition rate of a few minutes for peak fields near $\sim$30 Tesla. The system is complementary to split-pair magnets enabling one to perform various studies with magnetic fields lying in or very close to the scattering plane. The use of a separate cryostat for the sample allows precise positioning of samples in the bore while minimizing vibration propagating to the sample during pulsed-field generation. Finally, the system incorporates a double-funnel vacuum tube passing through the solenoid's bore in order to preserve the entire angular range allowed by the magnet. As such this system embodies a simple approach to preserve scattering angles for single-solenoid magnets providing 60 T and beyond that will become available in the forseeable future.

\section{Acknowledgment}
We have benefitted greatly from discussions with C. Swenson of LANL and from comments on the design of the system by  J. Schlueter, Y. Ren, and B. Brajuskovic of ANL. We thank J. C. Lang (ANL) for a critical reading of the manuscript. Use of the APS is supported by the DOE, Office of Science, under Contract No. DE-AC02-06CH11357. A part of this work was supported by International Collaboration Center at the Institute for Materials Research (ICC-IMR) at Tohoku University. HN acknowledges KAKENHI No.  23224009 from MEXT. JPCR acknowledges the support of NSERC of Canada. Work at Ames Laboratory was supported by the DOE, Office of Basic Energy Science, Division of Materials Sciences and Engineering. Ames Laboratory is operated for the DOE by Iowa State University under Contract No. DE-AC02-07CH11358.


\end{document}